\documentclass[apj, numberedappendix]{openjournal}

\usepackage[dvipsnames]{xcolor}
\usepackage[breaklinks,colorlinks,linkcolor=BlueViolet,citecolor=BlueViolet,urlcolor=BlueViolet]{hyperref}
\usepackage{amsmath}
\usepackage[autostyle]{csquotes} 
\usepackage[T1]{fontenc} 
\usepackage{orcidlink} 
\usepackage{needspace} 

\makeatletter
\renewcommand{\frontmatter@title@above}{%
  \vspace*{-\headsep}\vspace*{\headheight}%
  \footnotesize
  {\footnotesize The Open Journal of Astrophysics}\par 
  {\footnotesize\textsc{\@journalinfo}}\par 
  \vspace*{-\baselineskip}\vspace*{0.625in}%
}
\makeatother

\newcommand{\rhigh}{R_{\text{high}}}
\newcommand{\rlow}{R_{\text{low}}}

\newcommand{\spin}{a_{*}}

\newcommand{\PA}{\mathrm{PA}}
\newcommand{\paspin}{\mathrm{PA}_{\mathrm{spin}}}
\newcommand{\ispin}{i_{\mathrm{spin}}}
\newcommand{\pajet}{\mathrm{PA}_{\mathrm{jet}}}
\newcommand{\ijet}{i_{\mathrm{jet}}}
\newcommand{\uv}{(u,v)}
\newcommand{\quotes}[1]{``#1''}

\shorttitle{Position Angles in M87* and Sgr A*}
\shortauthors{Conroy et al.}

\begin{document}

\title{\vspace{-0.1cm}Ring Position Angles and Spin in M87* and Sgr A*\vspace{-1.5cm}}
	
\author{
	Nicholas S. Conroy\orcidlink{0000-0003-2886-2377},$^{1}$ 
	Michi Baub\"ock\orcidlink{0000-0002-5518-2812},$^{2,3}$ 
	Vadim Bernshteyn\orcidlink{0009-0000-1376-2352},$^{2,4}$
	Paul Tiede\orcidlink{0000-0003-3826-5648},$^{5,6}$\\
	Abhishek V. Joshi\orcidlink{0000-0002-2514-5965},$^{2,3}$
	Cora Prather\orcidlink{0000-0002-0393-7734},$^{6}$
	and Charles F. Gammie\orcidlink{0000-0001-7451-8935}$^{2,1,7,3}$
	\\
    {\scriptsize  \vspace{\baselineskip}
	$^{1}$Department of Astronomy, University of Illinois at Urbana-Champaign, 1002 West Green Street, Urbana, IL 61801, USA\\
	$^{2}$Department of Physics, University of Illinois at Urbana-Champaign, 1110 West Green Street, Urbana, IL 61801, USA\\ 
	$^{3}$Illinois Center for the Advanced Study of the Universe, University of Illinois at Urbana-Champaign,\\ 1110 West Green St., Urbana, IL 61801, USA\\
	$^{4}$Steward Observatory and Department of Astronomy, University of Arizona, 933 N. Cherry Ave., Tucson, AZ 85721, USA\\
	$^{5}$Center for Astrophysics $|$ Harvard \& Smithsonian, 60 Garden Street, Cambridge, MA 02138, USA\\
	$^{6}$Black Hole Initiative at Harvard University, 20 Garden Street, Cambridge, MA 02138, USA\\
	$^{7}$NCSA, University of Illinois at Urbana-Champaign, 1205 W. Clark St., Urbana, IL 61801, USA
    }
}

\begin{abstract}
Event Horizon Telescope (EHT) images of black holes appear as rings with a brightness asymmetry. 
Here, we expand on our previous study of the asymmetry magnitude $a_1$ to study the position angle of the peak brightness asymmetry $\PA_1$ in general relativistic magnetohydrodynamic (GRMHD) models.
For larger spin magnitudes ($\spin>0$ and $\spin\lesssim-0.5$), the mean $\PA_1$ falls within $1\sigma$ of the approaching limb of the black hole, regardless of viewing inclination, disk magnetization, or source.
By comparing the $(a_1, \PA_1)$ distribution in M87* observations with models, we demonstrate that we can mildly disfavor low-magnitude spins and strongly disfavor all spin vectors that point toward Earth.
The alignment of $\PA_1$ relative to the large-scale jet axis may suggest that M87*'s disk does not have a large tilt. 
By combining $\PA_1$ with the pattern speed measured in optimistic 2026 M87* video conditions, the EHT can constrain whether M87* is prograde or retrograde with $\sim 84\%$ accuracy. 
In Sgr A*, we show that a detection of $(a_1, \PA_1)$ could constrain the magnitude and direction of the galactic center spin vector.
Finally, if future EHT expansions increase the sample of horizon-scale sources, a simple set of observables (ring diameter, asymmetry magnitude, and asymmetry angle) could enable robust constraints on black hole mass, spin, and inclination.
\end{abstract}


\maketitle

\section{Introduction}\label{sec:intro}
The Event Horizon Telescope (EHT) has published the first images of the supermassive black holes M87* \citep{M87PaperI, M87_2018_paperI, M87_2021_paperI} and Sgr A* \citep{SgrAPaperI}. These images depict a ring of emission, produced by lensed near-horizon synchrotron radiation, with a brightness asymmetry.
The peak brightness asymmetry has a magnitude $a_1$ and a position angle $\PA_1$. In our convention, $(a_1, \PA_1)$ correspond to amplitude and phase of the fitted $m=1$ $m$-ring, and position angles are measured counterclockwise of North.

In earlier work \citep{Bernshteyn_2026}, we showed that $a_1$ depends on black hole spin $\spin\equiv Jc/GM^2$ (for black hole angular momentum $J$, speed of light $c$, gravitational constant $G$, and black hole mass $M$). 
The analysis used general relativistic magnetohydrodynamic (GRMHD) models, which have been crucial to the EHT theory work. Best-bet GRMHD models currently pass all observational constraints for M87* and correctly predicted the shift in $\PA_1$ from 2017 to 2018 \citep{M87PaperV, M87PaperVIII, M87PaperIX, M87_2018_paperI, M87_2018_paperII}. Using $a_1$ distributions from these models, we obtained a spin constraint for M87* with EHT intensity data. We found that $a_1$ measurements over three epochs of EHT M87* data were sufficient to marginally disfavor $|\spin |\lesssim 0.2$. This is consistent with the \cite{blandford_znajek} process powering M87*'s jet, in which magnetic fields threading the event horizon extract spin energy and power the jet.

\begin{figure*}
    \centering
    
    \includegraphics[width=\linewidth]{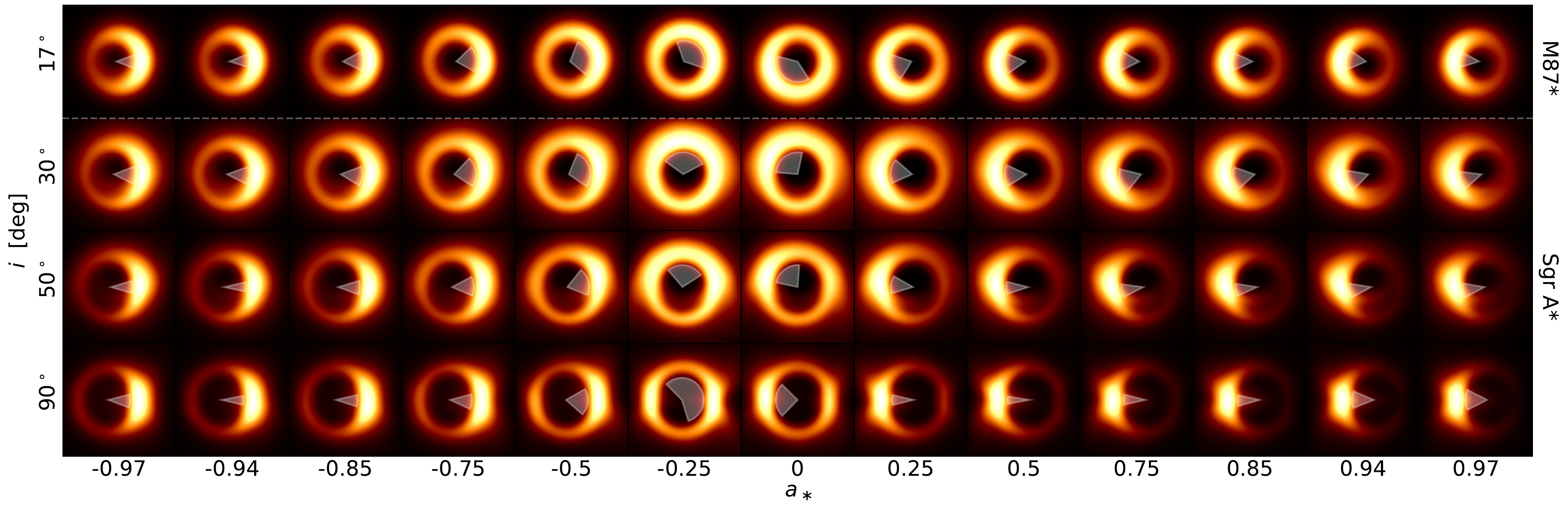}
   \caption{Mean MAD model images (averaging over time, $\rlow$, and $\rhigh$, and blurred using a $10\,\mu$as full width at half maximum Gaussian kernel) across spin and inclination, for both M87* ($i=17^\circ$) and Sgr A* ($i=30, 50, 90^\circ$). 
   White wedges shows the measured circular standard deviation $\sigma$ in $\PA_1$ about the circular mean.  Images are oriented so that disk angular momentum vectors project along $\PA=0$, prograde spin vectors project along $\PA=0$, and retrograde spin vectors project along $\PA=180^\circ$. 
   }
\label{fig:mean_pa} 
   \end{figure*}

In this work, we extend our prior analysis of $a_1$ to $\PA_1$. The brightness asymmetry is driven by a variety of factors, including Doppler boosting, gravitational lensing, frame dragging, radiative transfer effects, and gravitational redshift. Many of these factors depend on spin or exhibit correlations with spin, so like $a_1$, it is plausible for $\PA_1$ to also depend on spin. 
Figure \ref{fig:mean_pa} shows averaged images across a library of GRMHD models. These GRMHD models predict that $\PA_1$ typically falls on the approaching limb of the black hole, $90^\circ$ counterclockwise of the spin vector's projected position angle $\paspin$. 
While the typical dependence of $\PA_1$ on spin position angle has been known since \cite{M87PaperV}, there has been no dedicated study on the $\PA_1$ distribution or its dependence on model parameters. Here, we study the dependence of $\PA_1$ on spin, inclination, magnetization, and source.

After this Introduction, we review our GRMHD model library and our methodology for measuring $\PA_1$
with the Comrade and VIDA packages (Section \ref{sec:method}). We then analyze the dependence of $\PA_1$ on model parameters such as spin (Section \ref{sec:mean_pa}) and 
make predictions for $\PA_1$ dynamics (Section \ref{sec:dynamics}). By combining $\PA_1$ with other observables, we show that 2017-2018 data can constrain the M87* spin vector (Section \ref{sec:m87_constraints})
and that the upcoming 2026 M87* video may constrain whether M87* is prograde or retrograde (Section \ref{sec:m87_dynamics}). 
For upcoming EHT Sgr A* observations, we demonstrate $(a_1, \PA_1)$ could produce the first statistically significant constraint for the galactic center spin vector (Section \ref{sec:sgra}). We then discuss applications to other sources (which might be observed via EHT expansions such as the BHEX satellite; Section \ref{sec:future_sources}), implications for black hole growth histories (Section \ref{sec:growth_scenarios}), and limitations in our analysis (Section \ref{sec:uncertainties}). Finally, we  conclude (Section \ref{sec:conclusion}). 

\section{Measurement Technique}\label{sec:method}
\subsection{Model Library}
Here, we compare EHT data to a library of synthetic data. Following the PATOKA pipeline \citep{Wong_2022}, we use the \enquote{Illinois v5} library, which contains synthetic images of the \enquote{v5} version of the Illinois GRMHD models.
The library spans 5 parameters (magnetization mode MAD/SANE, black hole spin $\spin$, viewing inclination $i$, and electron temperature parameters  $\rhigh$ and $\rlow$), and is generated using 2 codes. 

First, we simulate the evolution of the accretion flow using KHARMA (\citealp{KHARMA_Prather_25}; based on \citealp{Gammie_03}; see \citealp{Wong_2022} for related details).
Each accretion disk falls into one of two magnetization modes: (1) standard and normal evolution \citep[\enquote{SANE};][]{sane1_2012, sane2_2013}, where the weaker magnetic field is not dynamically important, and (2) magnetically arrested disks \citep[\enquote{MAD};][]{mad1_1974, mad2_2003, mad3_2003}, where the magnetic field is dynamically important.

The dimensionless black hole spin is $\spin \equiv Jc/GM^2 \in [-1, 1]$. Prograde Kerr systems have $\spin > 0$, where the black hole spin and disk angular momentum vectors are parallel.
Retrograde systems have $\spin < 0$.
In a retrograde system with zero tilt, the black hole spin axis vector is antiparallel to the disk angular momentum vector. 
We span 13 values in black hole spin: $\spin \in \{0, \pm 0.25, \pm 0.5, \pm 0.75, \pm 0.85, \pm 0.94, \pm 0.97\}$ (N. J. Bowden et al., in preparation).

Next, we generate synthetic images using the radiative transfer code {\tt ipole} \citep{Mosc_2018}.  
The viewing inclination $i$ is defined as the angle between the observer line of sight and the angular momentum vector of the disk. 
In our convention, $i = 0$ is face-on above a counter-clockwise rotating accretion disk,  $i = 90^\circ$ is edge-on, and $i = 180^\circ$ is face-on below a clockwise rotating disk.
By default, the spin and disk angular momentum axes project onto the vertical axis of the image, unless the image $\PA$ is rotated to match observations.

For M87*, we consider $i \in \{17, 163^\circ\}$, assuming the angular momentum vector of the disk is coaxial with the large-scale jet \citep[see][]{Walker_2018}.
For Sgr A*, we currently span $i \in \{30, 50, 90^\circ\}$. Intermediate inclinations such as $i\sim 30^\circ$ are weakly favored by observations \citep{SgrAPaperV}.
We do not yet sample $i>90^\circ$ for Sgr A*, but
the distribution of $a_1$ is entirely independent of whether the disk is viewed above or below, while the $\PA_1$ distribution is reflected by symmetry.

The ion and electron temperatures may differ \citep[e.g.][]{Shapiro_1976}. The ion temperature $T_i$ is determined by KHARMA, while the electron temperature $T_e$ controls emission.
We assign the electron temperature during radiative transfer using the $\rhigh$ prescription \citep{rhigh_mosc_2016}: 
\begin{equation}\label{eq_edf}
    \frac{T_i}{T_e} = \rhigh \frac{\tilde{\beta}^2}{1+\tilde{\beta}^2} + \rlow \frac{1}{1+\tilde{\beta}^2}
\end{equation}
for $\tilde{\beta}=\beta/\beta_{\textrm{crit}}$, plasma $\beta \equiv P_{\textrm{gas}}/P_{\textrm{mag}}$, gas pressure $P_{\textrm{gas}}$, magnetic pressure $P_{\textrm{mag}}$, and $\beta_{\textrm{crit}}=1$.
Thus, $T_e$ smoothly varies from $\rlow$ when plasma $\beta \ll 1$ to $\rhigh$ when $\beta  \gg 1$. The parameter $\rhigh$ is the dimensionless maximum ratio of $T_i/T_e$. Physically, it corresponds to how efficiently we heat electrons in regions with low magnetic field strength (where there is typically less emission). 
For Sgr A*, we consider $\rhigh \in \{40, 160\}$ and $\rlow = 1$.
For M87*, we span $\rhigh \in \{10, 40, 80, 160\}$ and $\rlow\in \{1, 10\}$. In M87*, $\rhigh>1$ and $\rlow=10$ are motivated by observations \citep{M87PaperV, M87PaperVIII}. 

These models are imaged at a typical cadence of $5 \, GMc^{-3}$, corresponding to $\sim 100$ sec for Sgr A* and $\sim 44$ hr for M87*. They are imaged from $t= 20\,000 \, GMc^{-3}$ to $50\,000 \, GMc^{-3}$ for a total of $6\,000$ frames each. Hereafter, we refer to time in units of the gravitational time $t_g\equiv GMc^{-3}$. Images of our MAD models, averaging over time, $\rhigh$, and $\rlow$, can be found in Figure \ref{fig:mean_pa}.

\subsection{Asymmetry Measurements}

To measure ring asymmetry, we use the VIDA and Comrade packages \citep{tiede_vida_2022, tiede_comrade_2022}. 
VIDA fits a geometric model to an image using divergence minimization \citep[we use the default Bhattacharyya divergence;][]{tiede_vida_2022}. 
Comrade uses Bayesian inference to fit a geometric model to the raw visibility data.

Our geometric model is an $m$-ring truncated at $m=4$ with a Gaussian background component. The total intensity can be modeled as $I_{tot} = I_{ring} + I_{bg}$, for the ring intensity $I_{ring}$ and the background intensity $I_{bg}$. The Gaussian is:
\begin{equation}\label{eq_background}
I_{bg}(x, y) \propto (1- f) \exp\left(-\frac{(x')^2}{2\sigma_g^2}-\frac{(y')^2}{2\sigma_g^2(1+\tau_g)^2}\right)
\end{equation}
where $(1-f)$ scales the intensity of the Gaussian, $\sigma_g$ is the width, $\tau_g$ is the Gaussian ellipticity, $\xi_g$ is the rotation angle of the Gaussian, $(x_g,y_g)$ is the center, $x'=(x-x_g)\cos \xi_g +(y-y_g) \sin\xi_g$, and $y'=-(x-x_g)\cos \xi_g +(y-y_g) \sin\xi_g$.  
The $m=4$ $m$-ring model is: 
\begin{equation}\label{eq_ring_model}
I_{ring}(r, \theta) \propto \left[1 +  \sum_{m=1}^{4} a_m \cos \left( m \PA - \mathrm{PA}_m \right) \right] \exp \left( -\frac{(r-r_0)^2}{2\sigma_r^2} \right).
\end{equation} 
for ring radius $r_0$, ring thickness $\sigma_r$, brightness asymmetry magnitude $a_m$, position angle $\PA_m$, and Fourier component $m$. Notice this corresponds to a Fourier series truncated at $m=4$ and wrapped around a ring with Gaussian thickness.
This geometric model provides a good fit to both M87* and Sgr A* data \citep{M87_2018_paperI, M87_2021_paperI, SgrAPaperIV}. The fits are insensitive to minor changes in the geometric model or in the parameter bounds \citep[see Appendix B in][]{Bernshteyn_2026}.

The $m=1$ component captures the large-scale brightness asymmetry. Higher $m$ components typically have lower amplitudes, modifying the asymmetry shape or fitting small fluctuations. For this work, we consider $(a_1, \PA_1)$ as the overall brightness asymmetry magnitude and position angle respectively. 

We have applied the Comrade package to EHT 2017–2018 M87* observations, finding $(a_1, \PA_1)$ to be $(0.41 \pm 0.04, 169.7^\circ\pm 5.1^\circ)$ and $(0.69 \pm 0.10, 210.5^\circ\pm 3.5^\circ)$ respectively. These fits provide a good description of the EHT data \citep[Section 3.1]{Bernshteyn_2026}.\footnote{Unlike previous work \citep{Bernshteyn_2026}, here we only use 2017–2018 data, as 2021+ data remains proprietary to the EHT Collaboration.}

To simulate asymmetry data for M87*, we generate synthetic observations of the v5 M87* library. We rotate the ground truth images so each model's jet axis aligns with the observed jet axis 
\citep[the $-\vec{J}$ jet has position angle $\pajet=-72^\circ$ with standard deviation $<1^\circ$;][]{Walker_2018, M87_precessing_jet_1_Cui}, then Fourier transform the image and sample the visibilities assuming 2018 observational coverage and conditions. The results are consistent if we assume 2017 or 2021 observations \citep[see Appendix B of][]{Bernshteyn_2026}. Then we add observational noise \citep[including baseline-dependent thermal noise;][]{Chael_2018}. Finally, we optimize the parameters of our $m$-ring model using Comrade. To ensure the asymmetry samples are independent, we downsample to a cadence of $125\,t_g$, which is comparable to or longer than the correlation timescale. After downsampling, we have 240 asymmetry measurements for each combination of $(\spin,\rhigh,\rlow)$, and 1920 for each $\spin$.
Validation of the recovered $a_1$ values can be found in \cite{Bernshteyn_2026}. The effects of $\uv$ coverage on the recovered $\PA_1$ are described in Section \ref{sec:uncertainties}. 

For observed EHT Sgr A* data, extraction of $(a_1, \PA_1)$ is difficult, given limitations in snapshot $\uv$ coverage and the faster source variability of Sgr A*. Feasible methods include: 
(1)  static imaging, where a static image is generated then time-averaged image-domain features are measured via VIDA; 
(2) dynamical imaging, where a video is generated then time-variable image-domain features are measured via VIDA; 
(3) snapshot geometric modeling, where Comrade is applied to short time windows to measure time-variable features; and 
(4) full-track geometric modeling, where a geometric modeling pipeline fits source structure and source variability to measure time-averaged features \citep[]{SgrAPaperIII, SgrAPaperIV}.
In this work, we adopt the values reported using the third method in \cite{SgrAPaperIV}. That is, we adopt the $(a_1, \PA_1)$ values recovered from snapshot modeling using Comrade on April 6th and 7th, 2017. 

For simulated asymmetry data of Sgr A*, we use asymmetry measurements performed in the image domain via VIDA. As with synthetic M87* data, we downsample to a cadence of $125\,t_g$ to ensure asymmetry measurements in the Sgr A* models are independent.

\section{Results of Position Angle Analysis}\label{sec:results}
\cite{M87PaperV} was the first to report that the brightness asymmetry typically follows the black hole spin, rather than disk angular momentum. Here, we examine whether this finding endures under much denser spin sampling, spanning sources and inclination, and using state-of-the-art GRMHD models.

\needspace{2\baselineskip} \subsection{Mean Position Angle Morphology} \label{sec:mean_pa}
\begin{figure*}
    \centering
    \plotone{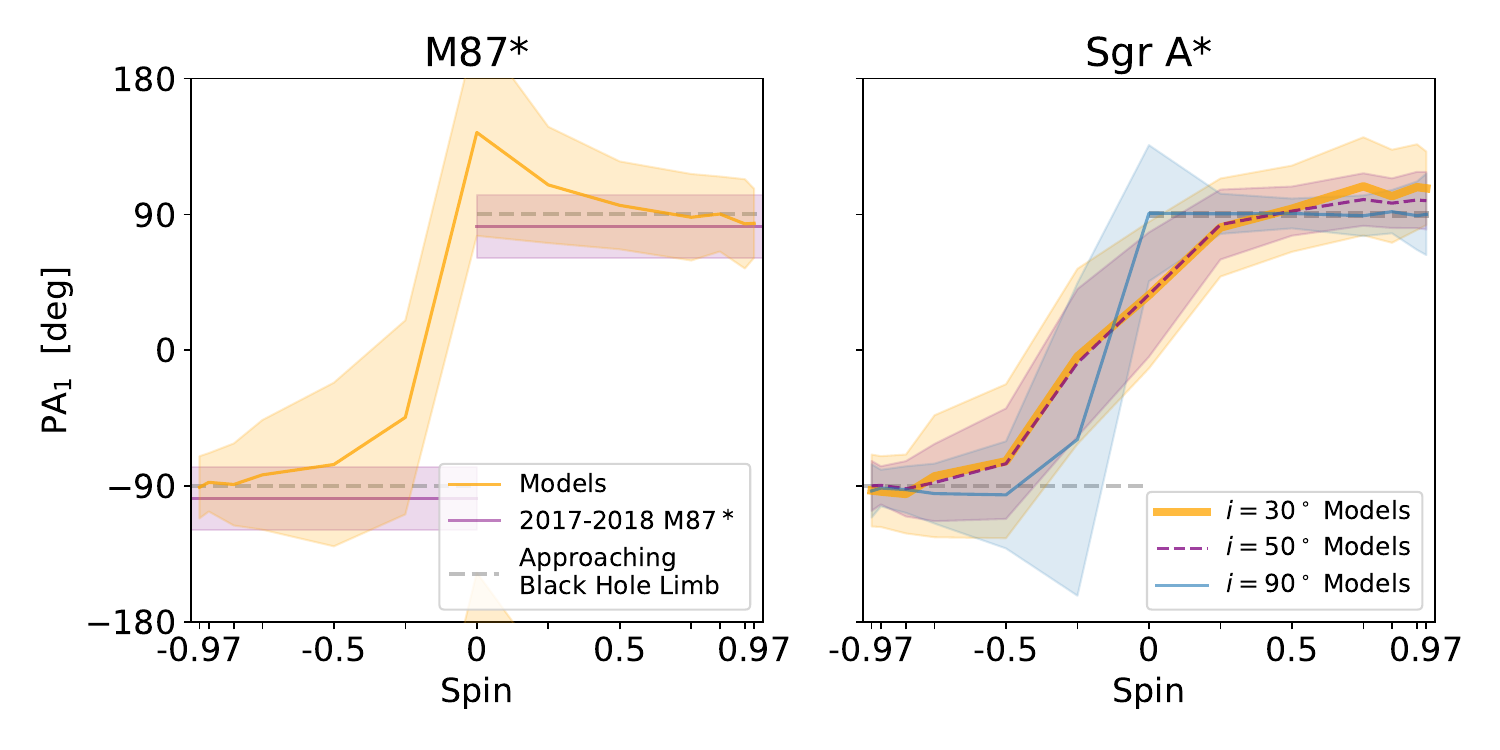}
   \caption{
   Distributions of $\PA_1$ for M87* (left) and Sgr A* (right). Lines correspond to the circular mean $\mu$ and colored bands correspond to the circular standard deviation $\sigma$ of the distribution. Images are oriented so that disk angular momentum vectors project along $\PA=0$, prograde spin vectors project along $\PA=0$, and retrograde spin vectors project along $\PA=180^\circ$. Across all high-spin models, $\PA_1$ falls within $1\sigma$ of the black hole's approaching limb. 
   }
\label{fig:gaussian_fit} 
   \end{figure*}
Figure \ref{fig:mean_pa} shows mean images for the MAD M87* and Sgr A*  models across all values of spin and inclination, averaging over time, $\rhigh$, and $\rlow$. For all models, the disk angular momentum vector projects upwards along $\PA=0$. In $a_*>0$ (prograde) models, the spin vector also projects along $\PA=0$, while in $a_*<0$ (retrograde) models, the spin vector projects along $\PA=180$. It is clear that for nearly all spinning models ($a_*> 0.25$ and $a_*\lesssim -0.5$), the brightness asymmetry falls along the black hole's approaching limb. 

This is demonstrated in Figure \ref{fig:gaussian_fit}, which plots the circular mean $\mu$ and circular standard deviation $\sigma$ of the $\PA_1$ distribution across spin. We find that $46/48$ of these spinning models have the mean $\PA_1$ fall within $1\sigma$ of the approaching limb. Exceptions occur for weakly retrograde Sgr A* models at low inclination ($\spin=-0.25$, $i=\{30^\circ, 50^\circ\}$). Here, spin-dependent effects (e.g. lensing) and disk-dependent effects (e.g. Doppler beaming) move the peak brightness to the opposite side of the ring. Unless a MAD model has a weakly retrograde spin, $\PA_1$ will fall on the side of the black hole's approaching limb.

Trends in $\sigma$ are also visible in Figures \ref{fig:mean_pa} and \ref{fig:gaussian_fit}. In particular, variations in $\PA_1$ seem to decrease roughly linearly with higher $a_1$. The asymmetry magnitude is minimized for zero-spin and weakly-retrograde models \citep{Bernshteyn_2026}; these models typically have larger $\sigma$ (see also Section \ref{sec:dynamics}).

There are slight differences in the $\PA_1$ distributions of M87* and Sgr A* near zero and weakly retrograde spins. One might suspect that differences in the methods contribute; Sgr A* asymmetry is measured in the image domain, while M87* asymmetry is measured in the $\uv$ domain (which relies on $\uv$ coverage that may be insensitive to large-scale extended structure). However, M87* $\PA_1$ measurements exhibit relatively little bias and dispersion from their image domain counterparts (see Section \ref{sec:uncertainties}). 
Instead, the differences are likely driven by changes in inclination and optical depth between the two sources. 
The more face-on inclination sampled for M87* influences which geodesics wrap through the emission region or are perpendicular to the magnetic field lines. Sgr A* is optically thicker, so its $\PA_1$ is sensitive to emission at a larger radius, where the effects of spin are diminished. Thus, as spin decreases from $0$ to $-0.5$, it takes longer for spin-dependent effects to dominate over disk-dependent effects in Sgr A* compared to M87*.  

Still, it is worth emphasizing the general consistency of the mean $\PA_1$, despite varying inclinations ($i=\{17^\circ, 30^\circ, 50^\circ,\allowbreak 90^\circ\}$) and varying sources (M87* and Sgr A*). These general trends are also consistent when varying the magnetization state (MAD vs SANE) and adiabatic index, as shown in Appendix \ref{sec:sanes}. Thus, across most model parameters, $\PA_1$ can be used for high-confidence constraint on $\paspin$:
\begin{equation}
\paspin \approx \PA_1-90^\circ
\end{equation}

\subsection{Expectations for Dynamics}\label{sec:dynamics}
For M87*, the EHT 2026 observing campaign will generate a time-lapse with a cadence of $\sim3.5$ days. For Sgr A*, improvements in $\uv$ coverage and imaging algorithms should also enable video reconstructions. 
It is thus interesting to consider $\PA_1$ dynamics.

\begin{figure*}
    \centering
    \plotone{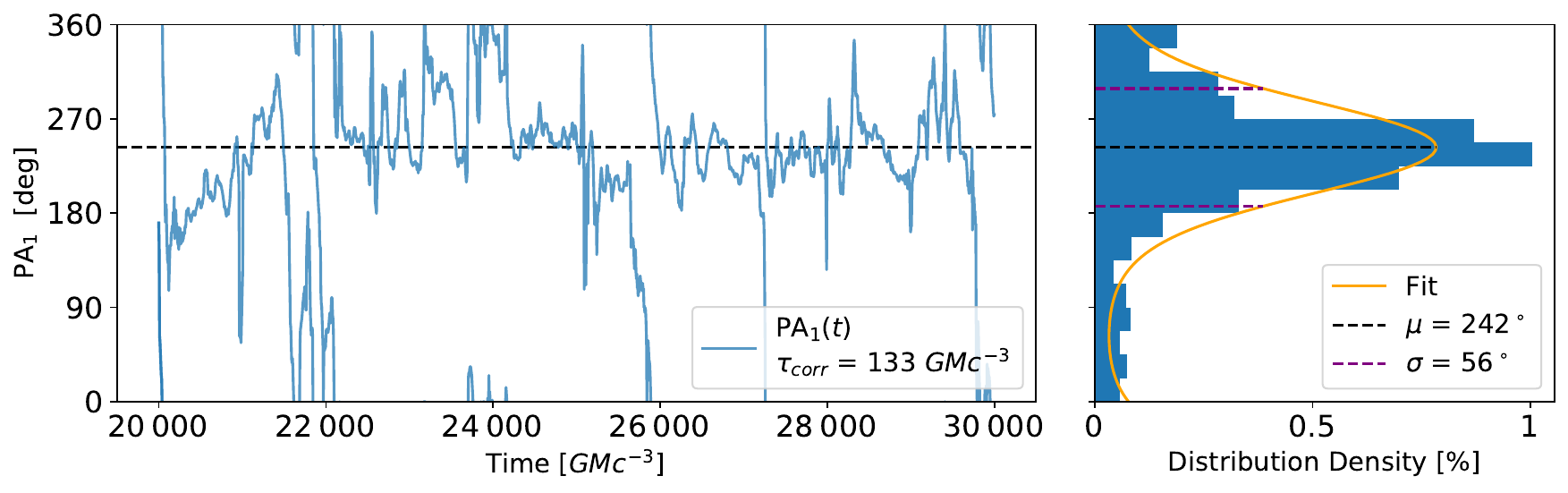}
   \caption{
   Timeseries of $\PA_1$ for a fiducial MAD spin-zero model (left) and the corresponding distribution density (right). The distribution evolves stochastically about the circular mean $\mu$ (dashed black line) with a circular standard deviation $\sigma$ (purple dashed line) and correlation time $\tau_{corr}$. Right, we fit a von Mises distribution (yellow line) as an approximation of the circular Gaussian distribution. 
   }
\label{fig:dynamics} 
   \end{figure*}

In Figure \ref{fig:dynamics}, we plot the time evolution of $\PA_1$ 
for a fiducial M87* model.
$\PA_1$ evolves stochastically around a mean value $\mu$. The right panel shows that $\PA_1(t)$ distributions are well described by a circular Gaussian distribution. The distribution also exhibits a correlation timescale $\tau_{corr}$. For this model, the correlation time of the position angle $\tau_{corr}(\PA_1) = 133\, t_g$ is longer than the correlation time of the asymmetry magnitude $\tau_{corr}(a_1) = 88\, t_g$.

Figure \ref{fig:dynamics} exhibits occasional windows of coherent $\PA_1$ evolution. Near $t\approx26000\,t_g$, we see $\PA_1$ complete a near full revolution around the ring. 
Coherent evolution sometimes corresponds to the motion of relatively bright features propagating around the ring, which drag the instantaneous $\PA_1$ away from the mean $\PA_1$.
Apparent rotation in the accretion flow typically follows $i$ \citep[i.e. the direction of the disk angular momentum vector relative to the line of sight, per][]{Conroy_2023, Conroy_2025}. Thus, like the pattern speed $\Omega_p$, these \enquote{$m$-ring dynamics} constrain the sign of $\cos i$. For example, the apparent rotation direction of the accretion flow might be extracted from a skew in the $d\PA_m/dt$ distribution. Analysis of $m$-ring dynamics could enable constraints on the near-horizon dynamics independent of imaging method, particularly relevant for future horizon-scale observations of new sources \citep{BHEX_2024}.

However, coherent swings in $\PA_1$ are intermittent. They require a sufficiently bright feature relative to the large-scale mean brightness asymmetry. Thus, as the mean brightness asymmetry magnitude increases, coherent $\PA_1$ swings become less frequent. $\Omega_p$ may offer a more direct, nonparametric measure for apparent near-horizon rotation than $m$-ring dynamics, if the imaging reconstructions are robust (see Section \ref{sec:m87_dynamics} for discussion of spin constraints using  $\PA_1$ and $\Omega_p$ ).

\section{Joint Constraints using Asymmetry Angle and Other Observables}\label{sec:results_magnitude}

\subsection{Spin Constraints with Static M87* Data} \label{sec:m87_constraints}

It is possible to constrain black hole spin using the brightness asymmetry. Here, we extend the methodology of \cite{Bernshteyn_2026} to form constraints with both the asymmetry magnitude and position angle $(a_1, \PA_1)$. 

In \cite{Bernshteyn_2026}, we applied the two-sample Kolmogorov-Smirnov (KS) test. 
It can test whether two samples (the asymmetry in observations and each model) come from the same underlying distribution, without assuming a parameterized form for the distribution.
It outputs the probability $p$ that the difference in the samples' cumulative distributions is as large or larger than measured, under the null hypothesis that the two samples are drawn from the same distribution.

\begin{figure*}
    \centering
    \plotone{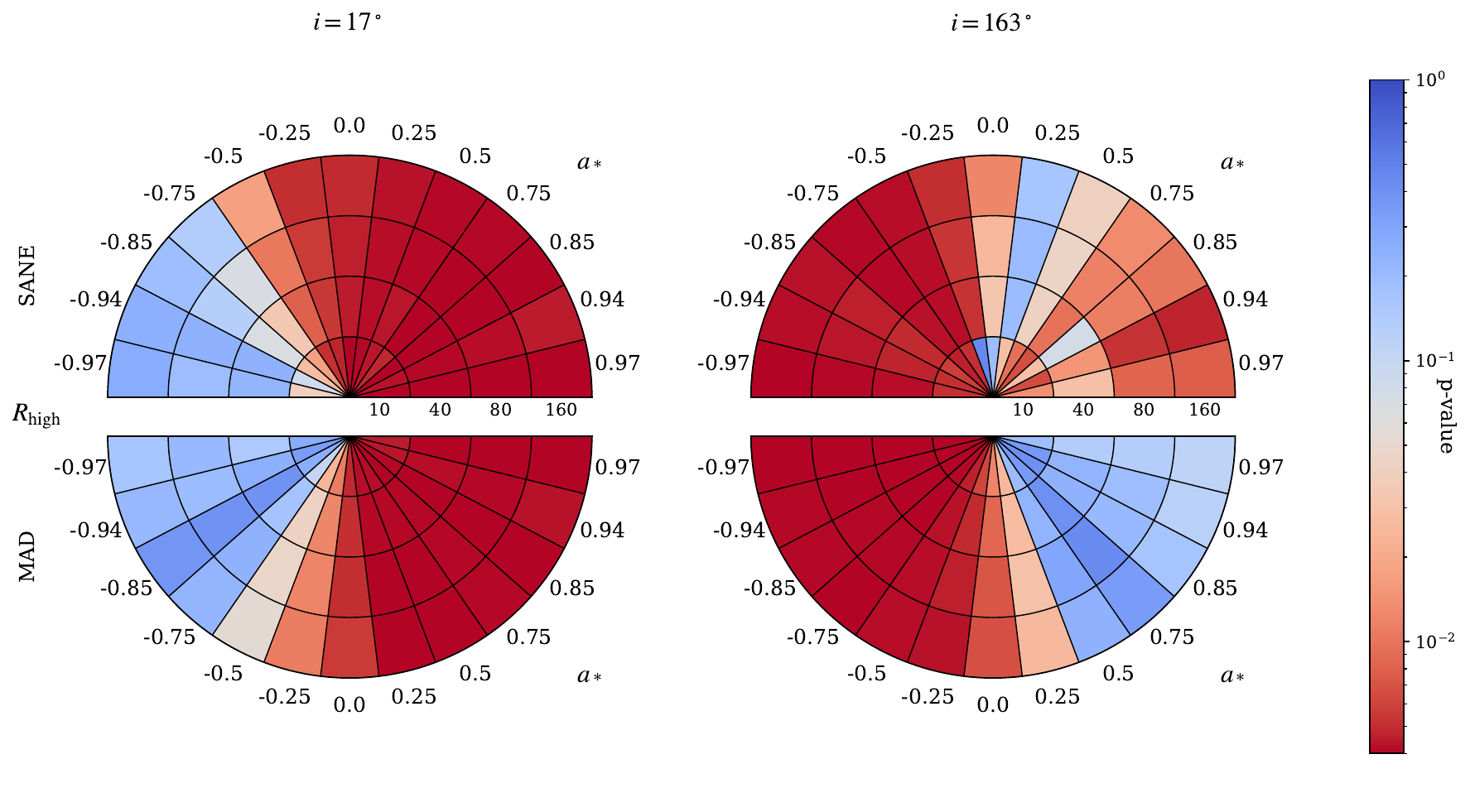}
   \caption{A \quotes{pizza plot} showing constraints based on the recovered asymmetry magnitude and position angle of M87* in 2017 and 2018. Subplot rows correspond to MAD and SANE, subplots columns correspond to inclination, radial position correspond to $\rhigh$ (marginalizing across $\rlow$), and azimuthal position correspond to spin. Models are disfavored when grey or red, for p-value $\lesssim 0.05$. Here, the asymmetry favors spinning black holes with a spin vector pointed away from Earth ($\spin \cos i < 0$). 
   }
\label{fig:pizza} 
   \end{figure*}

The KS test is only suited for one-dimensional, non-wrapped data. Thus we convert each asymmetry distribution $(a_1, \PA_1)$ to asymmetry phase space: $(a_{1x}, a_{1y})=(a_1 \cos\PA_1, \, a_1 \sin\PA_1)$. We then apply a standard implementation \citep{2D_KS} of the two dimensional KS test \citep{Peacock_83, Fasano_87, numericalrecipes} to the $(a_{1x}, a_{1y})$ data.

We illustrate the 2017–2018 M87* asymmetry constraints using a \quotes{pizza plot} (Figure \ref{fig:pizza}). We plot $p$ for each M87* model across model parameters, marginalizing across $\rlow$. Roughly half of each pizza is strongly disfavored with $p\ll0.05$. These low-$p$ wedges have $\spin \cos i >0$: their spin vector points towards Earth and so their $\PA_1$ lies in the North, which is strongly disfavored. While previous studies also found the M87* spin vector points away from Earth \citep{M87PaperV}, here we evaluate the claim statistically. Measurements of $\PA_1$ across two independent epochs favor M87* spin vectors pointed away from Earth with high probability.

For MAD models, Figure \ref{fig:pizza} also shows that low-magnitude spins are mildly disfavored. Zero-spin MADs have lower mean $a_1$  \citep{Bernshteyn_2026} and a wider $\PA_1$  dispersion (Section \ref{sec:mean_pa}) than observations. 
Having $|\spin|>0$ is consistent with predictions from the Blandford-Znajek model for jet-launching \citep{blandford_znajek}. Additional EHT data (2021+) would allow these low-spin MAD models to be more strongly ruled out 
\citep{Bernshteyn_2026}. 

For SANEs, larger $a_1$ variations do not yet allow us to entirely rule out low-magnitude spin using 2017-2018 $(a_1, \PA_1)$. Zero-spin SANEs, however, are disfavored by other observational constraints \citep{M87PaperV}. The asymmetry disfavors SANE, $i=163^\circ$, $\spin=0.94$ models more than neighboring spins at $\spin=0.85$ or $0.97$. Due to computational limitations, these SANE $\spin\in\{0.85,0.97\}$ models were imaged over a shorter duration ($t=20,000$–$30,000 \, t_g$), with a lower resolution, at a higher cadence (measuring asymmetry every $100\,t_g$ rather than every $125\,t_g$).
Thus, these SANE $\spin\in\{0.85,0.97\}$ models exhibit marginally larger variations in $a_1$ \citep{Bernshteyn_2026}. 
We predict that future observations (or consistent imaging methods for SANE $\spin\in\{0.85,0.97\}$ models) would allow us to disfavor this part of the parameter space, as highly prograde SANEs exhibit larger mean $a_1$ than observations \citep{Bernshteyn_2026}.

Favored model parameters are insensitive to methodology changes. For example, rather than applying a two-dimensional KS test to $(a_{1x}, a_{1y})$, we might apply two one-dimensional KS tests to $a_{1x}$ and $a_{1y}$ separately, or apply a KS test to $a_1$ and Kuiper's test (which allows for wrapped distributions) to $\PA_1$. The resulting $p$ values might be combined using methods such as Fisher's or Pearson's (which assume the $p$ are independent). These methods may change precise $p$ values, but do not affect favored or disfavored regions of the parameter space. Rather than calculating $p$, one might instead estimate the Bayesian evidence for each model. We leave the interesting problem of developing an optimal method of model–observation comparison to future work. 

Beyond spin constraints, $\PA_1$ may also enable constraints of disk tilt (where the disk angular momentum and spin vectors are not parallel or antiparallel).
Models predict the jet axis likely aligns with the disk angular momentum vector \citep{Liska_17_tilt}. Thus
if $\PA_1 \approx \paspin+90^\circ$ for non-zero tilt as well, then its location perpendicular to the jet axis would suggest that the disk and spin vectors are approximately coaxial, i.e. that M87*'s tilt is small. See Section \ref{sec:uncertainties} for additional tilt discussion.

Overall, current asymmetry measurements in M87* mildly favor a spinning black hole, and strongly favor spin vectors that point away from the Earth. Future EHT observations should improve these constraints.

\subsection{Spin Constraints with M87* Dynamics} \label{sec:m87_dynamics}

\begin{figure*}
    \centering
    \includegraphics[width=0.5\textwidth]{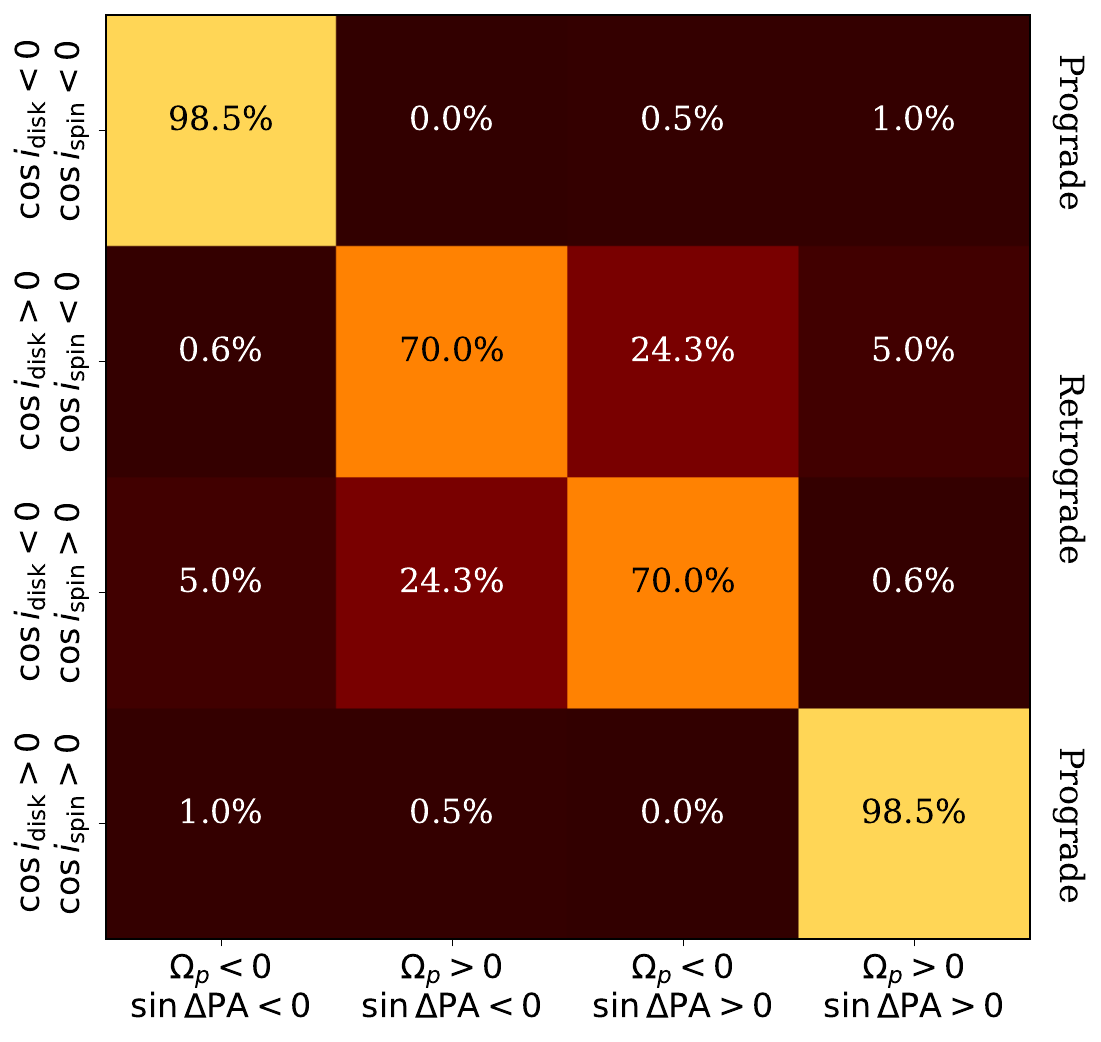}
   \caption{Constraints on the sign of M87* models' inclination and spin, based on model videos with realistic duration. We measure the direction of the pattern speed and whether $\PA_1$ occurs above or below the jet axis (i.e. the sign of $\sin \Delta \PA$ for $\Delta\PA \equiv \PA_1 - \pajet$). We recover the correct sign of the spin using $\cos i_{\mathrm{disk}}$ and $\cos \ispin$ $\sim84\%$ of the time (for inclination to the accretion disk angular momentum vector $i_{\mathrm{disk}}$ and inclination to the black hole spin vector $\ispin$).
   }
   \vspace{25pt}
\label{fig:spin} 
   \end{figure*}

For M87*, we have a prior for the disk and spin axes based on the approaching jet. Thus, we might constrain spin using $\PA_1$ and the pattern speed $\Omega_p$. Since $\PA_1$ depends on the direction of the black hole spin vector on the sky and $\Omega_p$ depends on the direction of the disk angular momentum vector on the sky, together they constrain whether the black hole is prograde or retrograde. 

In Figure \ref{fig:spin}, we show only the hemisphere of $\PA_1$  and the direction of $\Omega_p$ are needed to constrain $\spin>0$ vs $\spin <0$.
We divide each M87* model into time windows with 
duration $\Delta t=300\,t_g$.
The primary 2026 M87* observing campaign will have $\Delta t  \approx160\,t_g$, with $\Delta t  \approx310\,t_g$ including the dress rehearsal and possible extensions. We take three independent $\PA_1$ measurements under realistic observational conditions. We then test whether the circular mean $\PA_1$ falls above or below the approaching jet axis (i.e. whether $\sin \Delta \PA$ is positive or negative, for $\Delta \PA = \PA_1 - \pajet$ and approaching jet position angle $\pajet$). This depends on whether the spin vector aligns with the approaching or receding jet. We measure the sign of $\Omega_p$ in the image domain, without including observational noise or imaging error from limited $\uv$ coverage. This determines whether the disk angular momentum is approaching or receding.

How does this constrain the sign of the spin? 
The direction of a vector is completely described by its inclination and position angle: $(i,\, \PA)$. The approaching jet direction $(\ijet,\, \pajet)$ is well constrained. Assuming minimal tilt, EHT measurements of $\PA_1$ suggest the direction of the black hole spin vector $(\ispin,\, \paspin)$ aligns with the receding jet, i.e. that $\ispin\approx 163^\circ$. GRMHD models predict $\Omega_p \propto \cos i$, where $i=i_{\mathrm{disk}}$.
Thus, by constraining both $\ispin$ and $i_{\mathrm{disk}}$, we infer whether the spin and disk angular momentum vectors are aligned or antialigned. 

We successfully measure the disk and spin inclinations in $98.5\%$ of prograde samples, $70\%$ of retrograde samples, and $84.3\%$ across all samples. For retrograde models, we occasionally see pattern speeds rotating against the disk, partially from counter-rotation in the jet \citep[see][Figure 10]{Conroy_2025}. Retrograde spins also occasionally produce a brightness peak on the black hole's receding limb.
Two wrongs might make a right: if we measure the wrong sign for both $\cos i$ and $\cos \ispin$, we still recover the correct spin sign. In that case, we gain $99.5\%$ accuracy when prograde and $94.3 \%$ accuracy when retrograde. Our accuracy would worsen with imaging error for $\Omega_p$ and with shorter duration; it would  improve with additional observations \citep[e.g.,][]{Johnson_2023_ngeht, Midrange_science_goals, BHEX_2024} and other independent observables \citep[e.g.,][]{Palumbo_2020, Chael_2023, Ricarte_2025}.
Thus, the EHT 2026 observing campaign could constrain whether M87* is prograde or retrograde with $\sim 84\%$ accuracy. 

\subsection{The Spin Vector of Sgr A*} \label{sec:sgra}
There is currently no direct observational constraint on the spin vector of Sgr A*.  EHT asymmetry analysis has the potential to produce the first horizon-scale constraints on the galactic center spin vector. Here, we propose methods to infer Sgr A*'s spin and discuss what current 2017 data would suggest, while emphasizing that future observations with improved $\uv$ coverage are crucial for higher-confidence constraints.

\begin{figure}[t]
    \centering
    \includegraphics[width=0.5\textwidth]{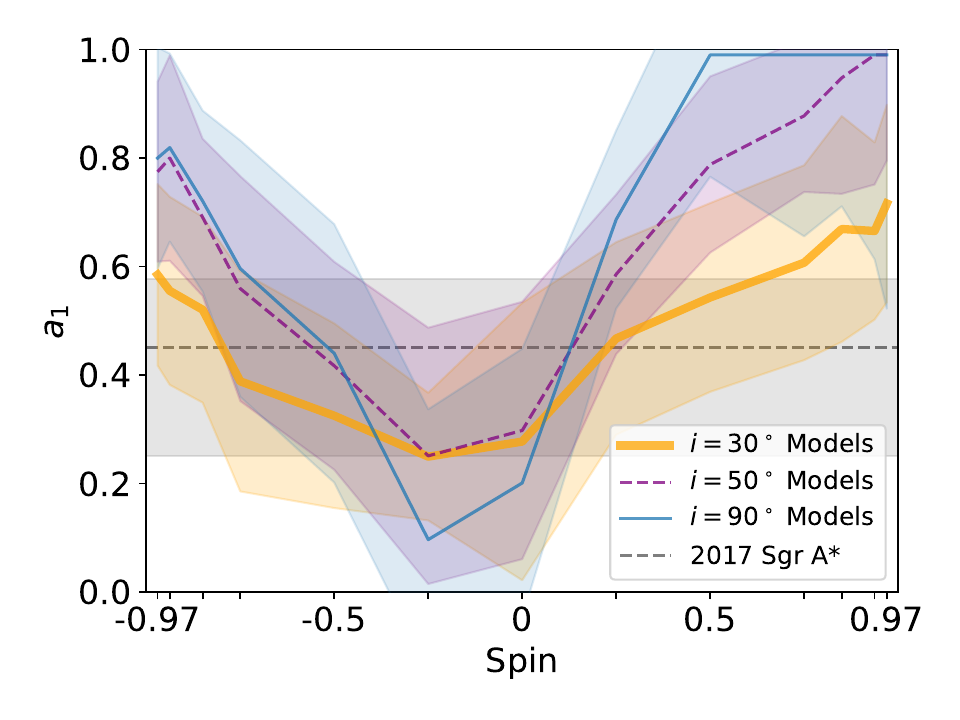}
    \caption{The asymmetry magnitude distribution in Sgr A* MAD models across spin at $i=30^\circ$ (yellow), $50^\circ$ (purple), and $90^\circ$ (blue). 
    Lines and surrounding shaded regions show the mode $\mu$ and standard deviation $\sigma$ respectively for the fitted truncated Normal distribution, without observational error from limited $\uv$ coverage.
    For observations, the dashed gray line shows the mean from snapshot geometric modeling with Comrade on April 6-7 2017; the surrounding gray $1\sigma$ shaded region shows observational uncertainty, based on uncertainty from Comrade and uncertainty across alternate pipelines added in quadrature \citep[see][Table 4]{SgrAPaperIV}.
    }
    \label{fig:spin_distr}
\end{figure}

Like M87* \citep{Bernshteyn_2026}, Sgr A* models exhibit a correlation between spin and brightness asymmetry magnitude $a_1$. In Figure \ref{fig:spin_distr}, we plot $a_1$ against spin, marginalizing across electron temperature parameters ($\rhigh, \rlow$) and time, for $i=30^\circ$, $50^\circ,$ and $90^\circ$. 
The $1\sigma$ regions of models and April 6-7 2017 observations overlap across all sampled values of spin for $i=30^\circ$. At higher inclinations, high $|\spin|$ models exhibit larger $a_1$ than current $1\sigma$ constraints. 
While the observational sample is small and observational error is not included in the model uncertainties, this suggests future observations could constrain Sgr A*'s spin magnitude using $a_1$ \citep[as in][]{Bernshteyn_2026}.

\begin{figure}[t]
    \centering
    \vspace{-10pt}
    \includegraphics[width=1\textwidth]{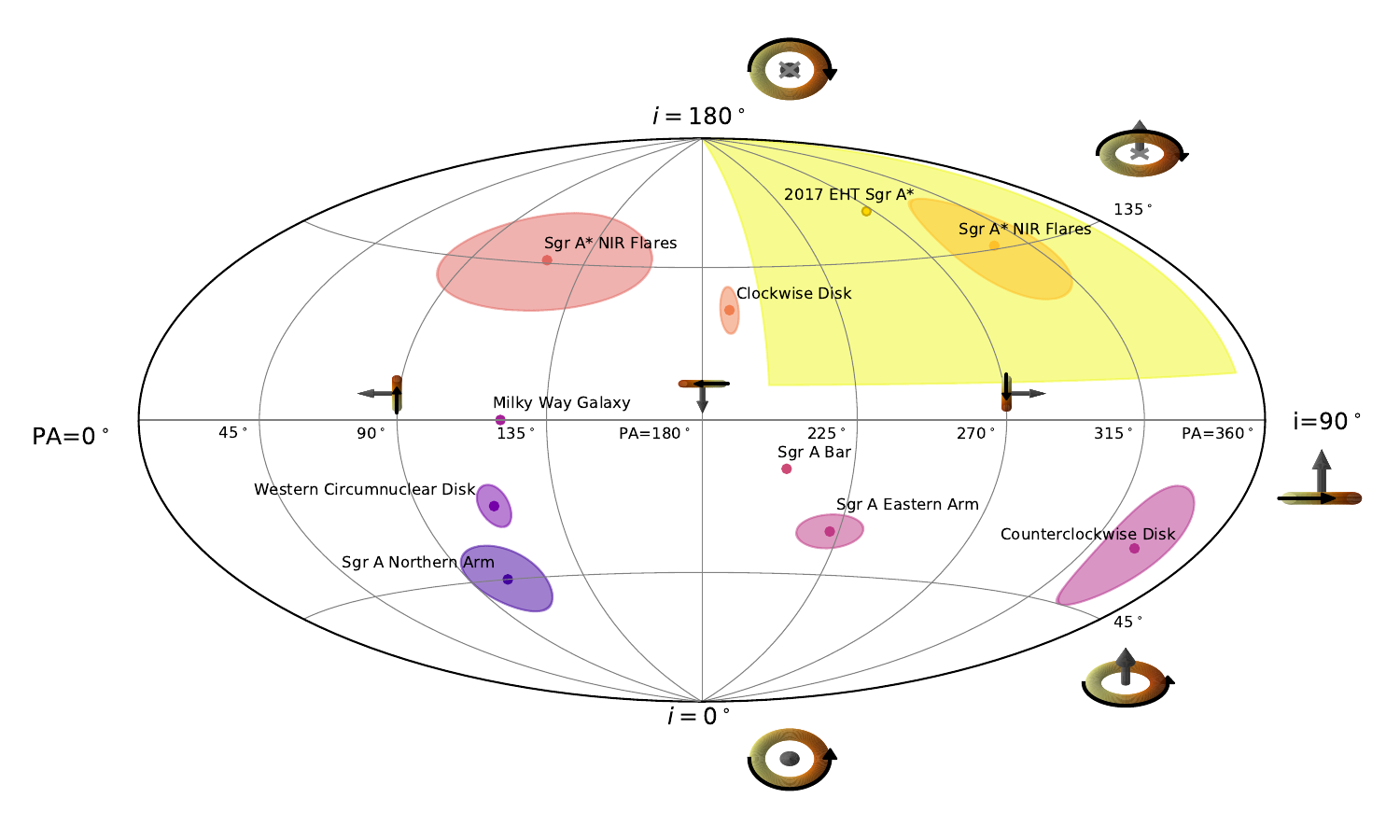}
    \vspace{-15pt} \caption{The direction of various galactic center $\vec{L}$ vectors as viewed from Earth, as a function of $\PA$ (horizontal axis) and $i$ (vertical axis). Shaded regions illustrate approximate $1\sigma$ uncertainties. Constraints on the Sgr A* spin vector (this work) using April 6-7 EHT data are shown in yellow, although we emphasize these constraints are tentative due to limited $\uv$ coverage. Other galactic center angular momentum constraints are colored based on the angular distance from Sgr A*. The Sgr A* spin vector appears most aligned with NIR flares and the clockwise stellar disk, potentially suggesting a correlation. See Footnote \ref{footnote} for additional details.
    }
    \label{fig:spin_diagram}
    \vspace{-5pt}
\end{figure}

What about the direction of the Sgr A* spin vector?
In Figure \ref{fig:spin_diagram}, we plot the direction of the inferred Sgr A* spin vector based on April 6-7 2017 observations relative to other galactic center 
structure.\footnote{
For 2017 EHT Sgr A*, $\paspin$ is estimated from Comrade $\PA_1$ measurements; the $1\sigma$ region includes Comrade uncertainties and the dispersion across other pipelines added in quadrature.
While the mean and $1\sigma$ uncertainties of $\ispin$ are difficult to estimate, EHT constraints favor $i_{\mathrm{disk}}>90^\circ$ \citep[due to linear and circular polarization;][]{Wielgus_2022, SgrAPaperVIII, Joshi_2024, Ricarte_2025}, disfavor edge-on, and mildly favor prograde spins near $\ispin=150^\circ$ \citep{SgrAPaperV, SgrAPaperVIII}. 
Thus, we take an approximate mean $\ispin=150^\circ$ and an approximate $1\sigma$ region $100^\circ \leq\ispin\leq 180^\circ$, emphasizing that future observations and improved parameter inference may shift these values.
Sgr A* NIR Flare constraints from GRAVITY Collaboration hotspot 
modeling
\citep{GRAVITY_2020_flares} have a degeneracy at $\PA \pm180^\circ$ due to uncertainty in when the inferred hotspot is approaching or receding. Other values are from  \cite{Genzel_2010_galactic_center}: structures include
the Sgr A Northern Arm \citep{Paumard_2004_galactic_center_northern_arm, Zhao_2009_galactic_center_northern_eastern_arm_cnd}, 
Eastern Arm \citep{Zhao_2009_galactic_center_northern_eastern_arm_cnd}, 
Bar \citep{Liszt_2003_galactic_center_bar}, 
Western arc of the Circumnuclear Disk \citep{Lacy_1991_galactic_center_cnd, Jackson_1993_galactic_center_cnd, Zhao_2009_galactic_center_northern_eastern_arm_cnd}, 
the Clockwise Disk \citep{Lu_2009_galactic_center_cw_disk, Bartko_2009_galactic_center_cw_disk}, 
and the Counterclockwise Disk \citep{Bartko_2010_galactic_center_ccw_disk}. 
Figure \ref{fig:spin_diagram} is inspired by \cite{Broderick_2016, Gravity_2023}.
\label{footnote}
}
The $1\sigma$ regions of our modest spin constraints overlap with
the GRAVITY collaboration's NIR centroid modeling of Sgr A* flares \citep{GRAVITY_2018, GRAVITY_2020_flares, Gravity_2023}. This may suggest relative alignment between Sgr A* and the near-horizon accretion disk based on mm and NIR data, consistent with expectations.
There is also relative alignment with the clockwise stellar disk, which contains Wolf–Rayet stars that feed the Sgr A* accretion disk \citep{Beloborodov_2006_galactic_center_cw_disk, Lu_2009_galactic_center_cw_disk, Ressler_2020_windfed, vonFellenberg_2022_galactic_center_cw_disk}. 

$\PA_1$ could also provide a novel constraint on the direction of the near-horizon Sgr A* jet. In M87*,  $\PA_1$ provides evidence for coaxiality between the spin vector and the large-scale jet axis. In Sgr A*, we might reverse this reasoning. Assuming sufficient coaxiality between the spin vector and the jet, $\PA_1$ could point the way to its jet. There is a long history of searching for the galactic center jet \citep[see][for a review]{Li_2013}. Current claims align in the following directions: 

(1) $\pajet \approx 60^\circ$,
which would predict $\PA_1 \approx -30^\circ$ or $150^\circ$
(claims include a near infrared feature per \citealp{Eckart_2006, Muzic_2007, Muzic_2010} and a radio feature per \citealp{Wardle_and_Yusef_92, Yusef_2012}); 

(2) $\pajet \approx 110^\circ$, which would predict $\PA_1 \approx 20^\circ$ or $200^\circ$ (claims include a radio feature per \citealp{Sofue_1989} and a gamma ray feature per \citealp{Su_and_fink_2012}); and 

(3) $\pajet \approx 135^\circ$, which would predict $\PA_1 \approx 45^\circ$ or $225^\circ$ (claims include a radio feature per \citealp{Yusef_1986}, a near infrared feature per \citealp{Eckart_2006}, and an X-ray feature per \citealp{Muno_2008}). 

EHT measurements of $\PA_1$ could provide an independent avenue for constraining the galactic center jet axis, but there are limitations. The constraints are model-dependent (see Section \ref{sec:uncertainties}).
Disk tilt would break the assumption of coaxiality. $\pajet$ may change across scales, if the jet is deflected by the surrounding medium. Improvements to short-baseline coverage may also enable direct imaging of the near-horizon jet.

Current data only enables tentative spin constraints in Sgr A*.
Estimates of $(a_1, \PA_1)$ are based on only two observations with limited  $\uv$ coverage. The effect of limited $\uv$ coverage on asymmetry extraction has not been fully understood. The optimal method for estimating $(a_1, \PA_1)$ with poor snapshot coverage is a topic of current research. There exists a large dispersion in $(a_1, \PA_1)$ across methods. Improved $\uv$ coverage and additional observations should enable more accurate estimates of brightness asymmetry and thus spin in Sgr A*. 

\section{Discussion}
\subsection{Future Sources Beyond M87* and Sgr A*} \label{sec:future_sources}
Future extensions to the EHT may enable horizon-scale asymmetry analysis for sources beyond M87* and Sgr A*. There are $\sim17$ sources with a ring angular diameter $\gtrsim5 \mu$as \citep{Pesce_2022, Ramakrishnan_2022_ETHER}. Ground-based multiwavelength extensions to the EHT \citep[see, e.g.,][]{AMT_Backes_2016, AMT_Labella_2023, ngEHT_2023_science, ngEHT_2023_array, Midrange_science_goals, ngEHT_2025_physics} could attempt to extract asymmetry in several new sources. The space-based mission Black Hole Explorer (BHEX) could provide horizon-scale analysis for 14 sources \citep{BHEX_2024, Zhang_2025, Akiyama_2026}. 

For these new sources, we propose \enquote{reading the ring:} using three total intensity ring observables that are simple to measure to enable straightforward parameter constraints.
(1) The ring angular diameter would constrain mass-to-distance ratio $M/D$ \citep[e.g.][]{M87PaperV}. (2) The brightness asymmetry magnitude would constrain the spin and inclination \citep{Bernshteyn_2026}. (3)  The brightness asymmetry position angle would constrain the spin vector position angle (this work). 

GRMHD models are now available for many potential future sources \citep{Zhang_2025}. Image reconstructions of these models demonstrate BHEX can successfully constrain the diameter, brightness asymmetry magnitude, and brightness asymmetry position angle of these sources \citep{Akiyama_2026}. This emphasizes the feasibility of ring morphology parameter inference.

Joint measurements of multiple observables would enable further parameter inference. 
If there is a prior for $\PA_{\mathrm{disk}}$ (e.g. from a large-scale jet),
then a joint measurement of $\Omega_p$ and $\PA_1$ would constrain whether the black hole is prograde or retrograde (as with M87* in Section \ref{sec:dynamics}).
A measure of $\PA_{\mathrm{disk}}$ and $\PA_1$
might also be used to infer disk tilt or low $|\spin|$.
For large tilts, $\PA_1$ may deviate from the expected value of $\PA_1 \approx \pajet \pm 90^\circ$ (see Section \ref{sec:uncertainties}).
$\PA_1$ in low $|\spin|$ models may also deviate from expectations (see Figure \ref{fig:mean_pa}). 
These scenarios could be distinguished using $a_1$, since low $|\spin|$ models exhibit the lowest $a_1$ \citep{Bernshteyn_2026}.
For nearly edge-on models, contributions from the $m=2$ mode $a_2$ begin to exceed $a_1$ (see Figure \ref{fig:mean_pa}). This suggests that if future observations can accurately recover $m=2$ $m$-ring parameters for new sources, the relative amplitudes of $a_1$ and $a_2$ may help identify whether the source is low-spin or edge-on. 

\subsection{Spin Alignment and Black Hole Growth History} \label{sec:growth_scenarios}
A joint constraint on $\spin$ and $\paspin$ could give insight into the black hole's growth history. 
Previous studies on growth history used cosmological simulations \citep[e.g.][]{Sala_2024_bh_growth_scenario, Beckmann_2025_bh_growth_scenario}, numerical accretion models, semi-analytical models \citep[e.g.][]{Ricarte_2025_spin_evolution}, and surrogate models \citep[e.g.][]{Wang_2024_bh_growth_scenario}. Three mechanisms are frequently proposed for supermassive black hole growth: (1) coherent accretion, where accretion roughly aligns with the galactic angular momentum vector; (2) incoherent accretion, where the accretion is uncorrelated with the galactic angular momentum vector; and (3) hierarchical mergers, where growth is driven by successive black hole–black hole mergers. Coherent accretion drives high $\spin$ that align with the large-scale gas flow. Incoherent accretion drives low $|\spin|$ that randomly misaligns with large-scale gas flow. Hierarchical mergers drive higher spins around $|\spin|\sim0.7$ \citep{Ricarte_2025_spin_evolution} that randomly misalign with the large-scale gas flow. 
Beyond growth histories, spin alignment also influences the orbits of nearby stars via Lense-Thirring precession \citep{levin_2003} and resonant friction \citep{levin_2024}.

In M87*, $a_1$ disfavors $|\spin| \lesssim 0.25$ and mildly favors $|\spin|\gtrsim 0.5$. 
$\PA_1$ suggests the spin vector is relatively aligned with the large scale jet, and so is tilted from angular momentum of the surrounding pc-scale gas environment \citep[recent studies suggest the tilt between the pc-scale gas and the jet is $\beta \gtrsim11^\circ$, with a more typical value of $\beta\approx27^\circ$, per][]{Jeter_2021}. 
Non-zero spin disfavors incoherent accretion. Sufficiently large tilt from the pc-scale gas may suggest mergers played a role or that the large-scale gas has changed orientation. Precise constraints on the growth history will depend on the spin and the model. 

\needspace{4\baselineskip} \subsection{Uncertainty in Position Angles and Spin Constraints} \label{sec:uncertainties}
The analysis presented here is model-dependent. 
Going sequentially, there are three sources of uncertainty in the GRMHD simulations that might affect $\PA_1$.

(1) Boundary and initial conditions. Our models are typically initialized with an aligned Fishbone-Moncrief torus, with magnetic field conditions that produce the MAD or SANE state, and without inflow at the outer boundaries \citep{Fishbone_moncrief_torus, Wong_2022}. 

The v5 library also contains several tilted disks. We consider $\sim30$ independent samples from a set of Sgr A* models with $a_*=\pm0.5$, tilt $\beta=10^\circ$, $\rhigh\in\{40, 160\}$, viewed at $\ispin=30^\circ$ with respect to the spin axis and at azimuth $\alpha\in\{0^\circ, 90^\circ\}$ (where $\alpha=0^\circ$ corresponds to a disk tilted out of the screen towards the observer, and $\alpha=90^\circ$ corresponds to a disk tilted within the plane of the screen). All tilted models exhibited a mean $\PA_1$ within $1\sigma$ of the black hole's approaching limb and the untilted mean $\PA_1$. This is consistent with previous studies of mild tilts. Deviations in $\PA_1$ from expectations may increase with sufficiently large tilt ($\beta \gtrsim 30^\circ $) for certain $\spin$, $i$, and $\alpha$ \citep{Chatterjee_2020, M87_2018_paperII}. 

SANE models are discussed in Appendix \ref{sec:sanes}, showing broadly consistent trends. The effect of alternate boundary conditions (such as wind-fed GRMHD models) on $\PA_1$ is not yet understood.

(2) Collisionless dynamics. Our simulations assume the plasma is an ideal fluid, and thus do not include kinetic or non-ideal fluid effects. Recent 2D global kinetic simulations, which include kinetic effects (and a more complete electron energization treatment), produce remarkably similar results to MAD GRMHD models \citep{Mehlhaff_2026_globalPIC}. Non-ideal GRMHD modeling with viscosity and heat conduction produce nearly indistinguishable mean asymmetry behavior \citep[see Figures 3 and 9 of][]{Dhruv_nonideal_2025}.

(3) Thermodynamics and electron energization. In our models, the ion internal energy is evolved assuming a set adiabatic index $\gamma_i$,  with $\gamma_i = 4/3$ and $\gamma_i = 5/3$ in the v3 and v5 libraries respectively. The larger adiabatic index is favored for current EHT sources \citep{Gammie_2025}. Still, Appendix \ref{sec:sanes} shows that results are consistent for lower $\gamma_i$. 

Our models assume a thermal electron distribution function, following the standard $\rhigh$ prescription \citep{rhigh_mosc_2016}. One sample of GRMHD models with a nonthermal prescription exhibit surprisingly similar mean values of $(a_1, \PA_1)$ \citep[see][Figure 3]{M87_2018_paperII}. A more accurate \enquote{two-temperature} treatment would evolve a separate electron internal energy equation, with electron heating, cooling, and a nonthermal tail \citep{HongXuan_2023, Salas_2025}. One two-temperature GRMHD library shows starkly similar $\PA_1$ behavior, with $\PA_1$ falling on the black hole's approaching limb for all $|\spin|\gtrsim 0.3$, wider standard deviations for $|\spin|\lesssim 0.3$, and a bias towards larger $\PA_1$ than the approaching limb for $|\spin|\lesssim 0.3$.
\citep[compare Figure \ref{fig:gaussian_fit} here with Figure 11 of][]{Chael_2025}.

We assume the electron temperature is isotropic.
Increased temperature anisotropy perpendicular to the magnetic field in the mirror limit or whistler limit produces similar $\PA_1$ across multiple sources, inclinations, and spins. Increased anisotropy parallel to the magnetic field with the firehose limit can change the mean $\PA_1$ if $\spin$ is low \citep{Galishnikova_23_anisotropy}.
However, these limits are disfavored by the observed circular polarization fraction \citep{ginzburg_1965, Galishnikova_23_anisotropy, M87PaperIX} and by 2D global kinetic simulations \citep[i.e. by particle-in-cell simulations;][]{Mehlhaff_2026_globalPIC}. 

There are three additional uncertainties in the radiative transfer treatment:

(4) Mass density scaling. While the equations of GRMHD are scale invariant, radiative transfer requires scaled quantities in order to calculate outputs in physical units. The accretion mass density $\mathcal{M}$ is thus scaled to match the observed source's flux density. There may be uncertainties in the source's target flux (from annual variability or unresolved large-scale emission), and future sources may have different values of $M$, $D$, and $\mathcal{M}$. Large changes in these parameters affect optical depth and thus source morphology \citep{Wong_2022}. We leave an analysis of $(a_1, \PA_1)$ across changes in $M$, $D$, and $\mathcal{M}$ to future work. 

(5) The \enquote{fast-light} approximation. We perform radiative transfer for a snapshot image along a single GRMHD timeslice, i.e. ignoring changes in the source during the light-crossing time. \cite{Bernshteyn_2026} demonstrated that, for a fiducial MAD model, this fast-light approximation has no measurable effect on $a_1$. Using the same model, we find $\PA_1$ is also unchanged. The fast-light treatment generates a mean and standard deviation of $\PA_1=191^\circ \pm22^\circ$, while slow-light produces $\PA_1=194^\circ \pm22^\circ$.

(6) Inclination uncertainty, in the case of M87*. Our models of M87* assume $i=17^\circ$ or $163^\circ$, based on the inclination of the large-scale jet \citep[which has a $1\sigma$ uncertainty of $\sim3^\circ$ to $4^\circ$;][]{Mertens_jet_2016, Walker_2018}. We consider four MAD models, with $\spin=\pm0.5$ and with a shift of $\pm5^\circ$ from the standard inclination. The shift from more edge-on to more face-on did not shift the mean $\PA_1$ (within $\pm0.1^\circ$), but did increase the standard deviation (by $\sim80\%$), both as expected.

Finally, there are two sources of uncertainty when measuring $\PA_1$ in observations and inferring $\paspin$.

\begin{figure*}
    \centering
    \plotone{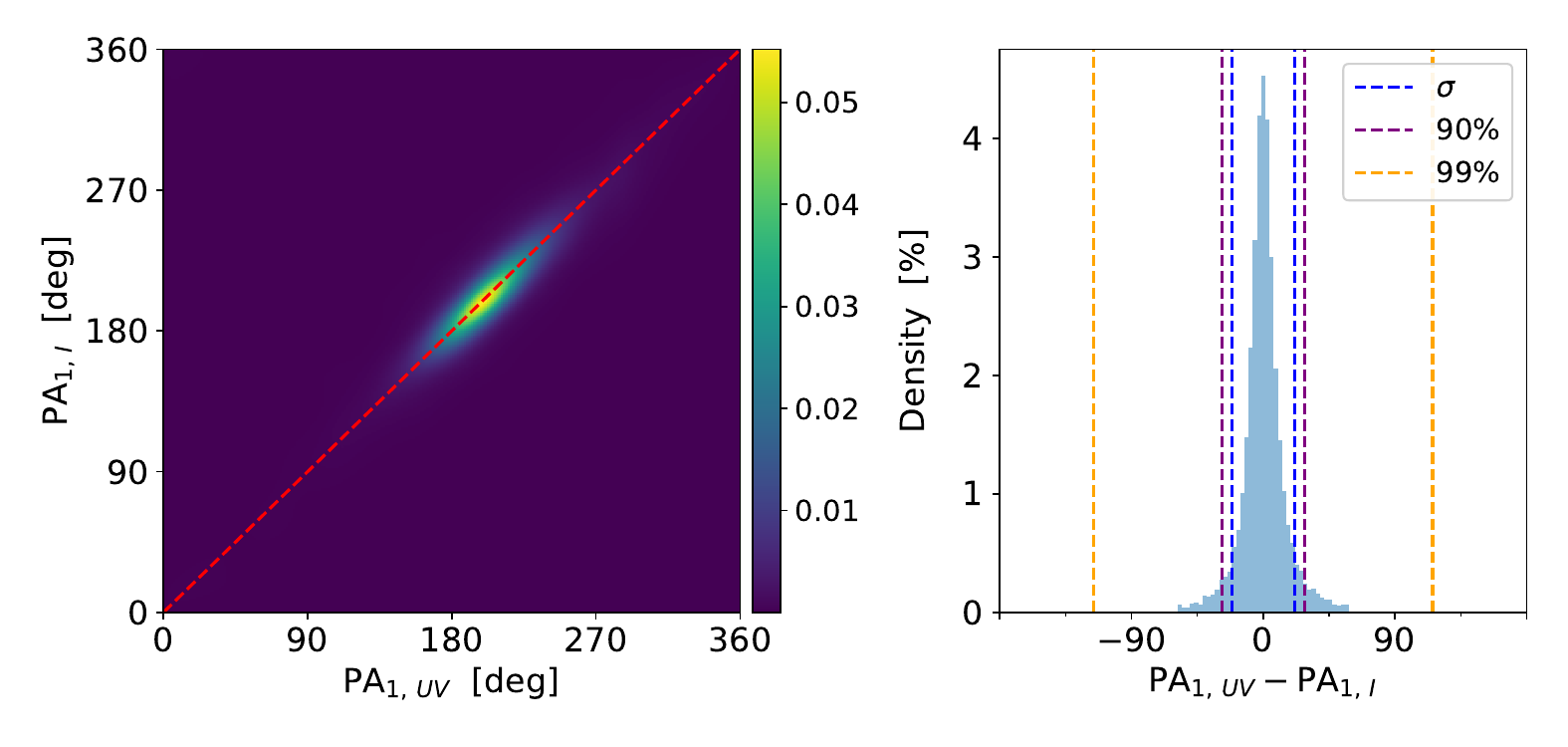}
   \caption{Left: the density of the $m=1$ position angles for M87* MAD models measured in the image domain ($\PA_{1,I}$) and
   $\uv$ domain ($\PA_{1,UV}$) assuming 2018 M87* $\uv$ coverage, calculated using Gaussian kernel density estimation. The dashed red line corresponds to $\PA_{1,UV}=\PA_{1,I}$. Right: The density of position angle measurements across $\PA_{1,UV}-\PA_{1,I}$, with $1\sigma$ (for circular standard deviation $\sigma$), $90\%$, and $99\%$ boundaries marked as dashed lines. $90\%$ of snapshots are within $28^\circ$.
   }
\label{fig:uv_vs_im} 
   \end{figure*}

(7) Measurement error. In Figure \ref{fig:uv_vs_im}, we show the effects of limited $\uv$ coverage. We plot $\PA_1$ measurements in the image domain ($\PA_{1,I}$) and in the visibility domain with 2018 M87* observing conditions ($\PA_{1,UV}$). The $1\sigma$ region encloses all snapshots within $\pm21^\circ$; $90\%$ of all snapshots are within $\pm28^\circ$. Notice that the distribution over $\PA_{1,I} - \PA_{1,UV}$ is unbiased. The $1\sigma$ region is not significantly larger than the standard deviation of the blurring kernel due to EHT resolution. Further, it is smaller than the standard deviation from variations within a single model. With M87* $\uv$ coverage, it seems we are able to accurately recover $\PA_1$. As snapshot $\uv$ coverage worsens for other sources, we expect the error to increase.

(8) Statistical error. If $\PA_1$ is normally distributed, then uncertainty in the measured mean scales as $\sigma/ \sqrt N$ while the uncertainty in the measured standard deviation scales as approximately $\sigma / \sqrt {2N}$, for true distribution standard deviation $\sigma$ and $N$ independent samples. The mean $\PA_1$ may not precisely align with the black hole's approaching limb, potentially introducing a bias in the inferred $\paspin$. Model predictions for $\sigma$ and $\PA_1$ can be found in Figure \ref{fig:gaussian_fit} and Appendix \ref{subsec:appendix_data}. All highly spinning models produce a mean $\PA_1$ within $1\sigma$ of the black hole's approaching limb.

\section{Conclusions}\label{sec:conclusion}
Here, we have analyzed the position angle of the brightness asymmetry seen in EHT models and sources. The direction of the brightness asymmetry depends on spin: $\PA_1=\PA_1(\spin)$. Thus, a measurement of $\PA_1$ corresponds to a constraint  of the black hole spin vector on the sky.

In M87*, we show a joint measurement of the brightness asymmetry magnitude and direction $(a_1, \PA_1)$ allows us to disfavor low spins. This is discussed in detail in prior work \citep{Bernshteyn_2026}.
Further, we can disfavor high spins pointed toward Earth, providing a statistically significant model-dependent constraint that M87* has $|\spin| \gtrsim0.2$ and has a spin vector pointed along the receding jet. The alignment of $\PA_1$ perpendicular to the jet axis may allow us to infer a limited tilt in the M87* accretion disk.
The 2026 M87* video should provide a joint constraint on the pattern speed $\Omega_p$ and $\PA_1$, constraining whether M87* is prograde or retrograde with $\sim84\%$ confidence.

In Sgr A*, we demonstrate that EHT data is beginning to constrain $(a_1, \PA_1)$ with high uncertainty. Future analysis should enable robust horizon-scale constraints on the direction and magnitude of the Sgr A* spin vector. This could offer insight into its growth history, point the way to the near-horizon jet, and contextualize studies of galactic center stellar orbits.

Future expansions to the EHT such as BHEX may enable us to perform horizon-scale analysis on over a dozen sources. A set of simple total intensity ring observables should enable straightforward parameter inference in these new sources. The ring diameter is known to scale with $M/D$. The brightness asymmetry magnitude scales with spin magnitude and inclination \citep{Bernshteyn_2026}. And the brightness asymmetry position angle depends on the spin vector position angle. This will complement independent constraints through other observables, such as the linear polarization fraction and structure.

Spin constraints using brightness asymmetry could offer insight into jet-launching, disk tilt, and potentially black hole growth history. 
The leading model for jet-launching is Blandford-Znajek, which predicts that jet power is a function of spin. 
Prior M87* analysis of $a_1$ enables spin constraints that are consistent with Blandford-Znajek \citep{Bernshteyn_2026}. Analysis of $\PA_1$ here is consistent with our prior results.
Improved constraints using future independent observations will enable an indirect test for Blandford-Znajek, by checking continued consistency between spin and jet power. 
$\PA_1$ provides independent evidence that the M87* spin vector is relatively aligned with the jet axis, i.e. that the disk has little tilt, consistent with prior results \citep{M87_precessing_jet_1_Cui}. If Sgr A* also exhibits little to no tilt, $\PA_1$ may point the way to the near-horizon jet. 
In other sources, $\PA_1$ may constrain whether the black hole spin vector is aligned with or tilted from the disk and larger galactic center structure. 
More generally, joint measurements of $(a_1, \PA_1)$ may offer  insight into black hole growth, as different growth histories predict different probability distributions for black hole spin magnitudes and spin alignment with the surrounding galactic structure.

\needspace{2\baselineskip} \section*{Acknowledgments}

We thank Angelo Ricarte for helpful discussions, and for comments that improved this work. We thank George Wong for providing access to the slow-light GRMHD model. Finally, we thank the anonymous referee for their comments which also improved this work. This work was typeset using \LaTeX\ style openjournal v. 09/06/15.
This work was supported by the NSF grant AST  20-34306. This material is based upon work supported by the National Science Foundation under Grant No. PHY-2244433. 
N.C. is supported by the NASA Future Investigators in NASA Earth and Space Science and Technology (FINESST) program. This material is based upon work supported by the National Aeronautics and Space Administration under Grant No. 80NSSC24K1475 issued through the Science Mission Directorate.
This research used resources of the Oak Ridge Leadership Computing Facility at the Oak Ridge National Laboratory, which is supported by the Office of Science of the U.S. Department of Energy under Contract No. DE-AC05-00OR22725.  This research used resources of the Argonne Leadership Computing Facility, which is a DOE Office of Science User Facility supported under Contract DE-AC02-06CH11357. This research was done using services provided by the OSG Consortium, which is supported by the National Science Foundation awards \#2030508 and \#1836650. This research is part of the Delta research computing project, which is supported by the National Science Foundation (award OCI 2005572), and the State of Illinois. Delta is a joint effort of the University of Illinois at Urbana-Champaign and its National Center for Supercomputing Applications. We are particularly grateful to the Argonne Leadership Computing Facility for providing storage space on the Eagle system that was critical for enabling this work.  This work was initiated in part at the Aspen Center for Physics, which is supported by National Science Foundation grant PHY-2210452. 

\begin{appendix}
\section{SANE Models and the v3 Illinois GRMHD Library}\label{sec:sanes}
\begin{figure}
    \centering
    \includegraphics[width=0.6\textwidth]{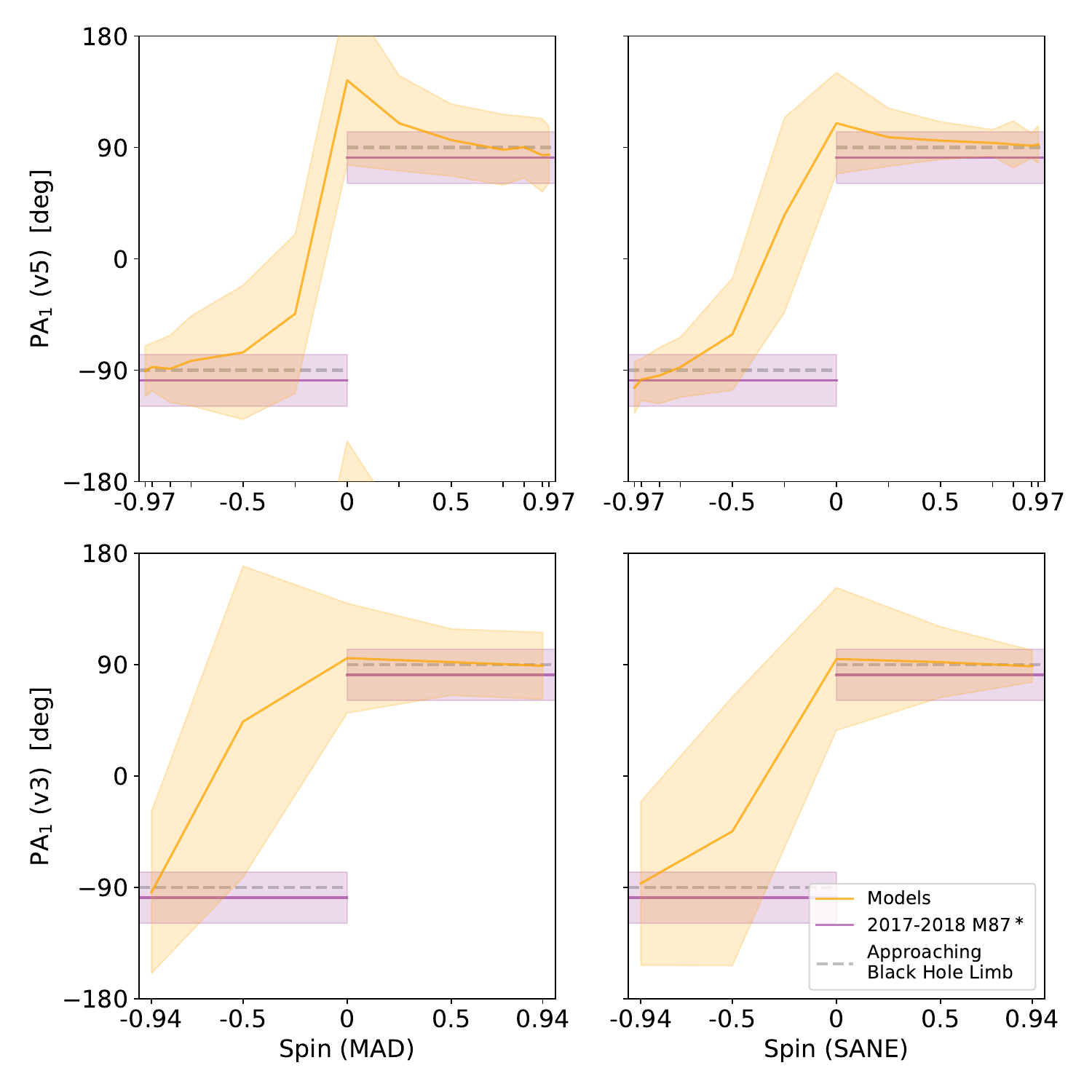}
    \caption{As in Figure \ref{fig:spin_distr}, we plot $\PA_1$ circular means with $1\sigma$ ranges, now for MAD (left subplots), SANE (right), v3 (bottom), and v5 (top) models.
    Models are shown in orange, 2017-2018 M87* is shown in purple, and the black hole's approaching limb is shown as a dashed gray line.  X-axis ticks show the spins sampled in each model library.  Images are oriented so that disk angular momentum vectors project along $\PA=0$, prograde spin vectors project along $\PA=0$, and retrograde spin vectors project along $\PA=180^\circ$. Consistent trends in $\PA_1$ across magnetization and library versions are visible.   }
    \label{fig:model_comparison}
\end{figure}

The majority of the above analysis focused on MAD models from the latest v5 model library. In Figure \ref{fig:model_comparison}, we show the $\PA_1$ distributions for weakly magnetized SANE models and the earlier v3 version of the Illinois GRMHD model library. All models show broadly consistent trends: $\PA_1$ falls within $1\sigma$ of the black hole's approaching limb for all models with $\spin>0$ or $\spin\lesssim-0.5$ 

In SANE models, the rotation curve is roughly Keplerian, unlike MAD models which exhibit sub-Keplerian rotation \citep{Conroy_2023, Dhruv_grmhd_survey_2024}. Thus, Doppler beaming plays a larger role in driving asymmetry in SANEs.
This generates larger $a_1$ for prograde spins and shifts the minimum $a_1$ to occur at more retrograde spins \citep{Bernshteyn_2026}. Nevertheless, the mean $\PA_1$ exhibits surprisingly similar behavior between MADs and SANEs, suggesting our results are relatively insensitive to the accretion magnetization.

Compared to the newer v5 library, the v3 library was run for a shorter duration ($t=15\,000$–$30\,000 \, t_g$) with less resolution in the grid, sampling different $\rhigh$ values ($1$ instead of $80$) and fewer spins ($\spin\in\{0, \pm0.5, \pm0.94\}$). It also used an ion adiabatic index of $\gamma_i = 4/3$ instead of $\gamma_i = 5/3$, which is now preferred \citep{Gammie_2025}. The effect of each of these changes has not yet been fully investigated. 

$\PA_1$ generally shows similar trends between the two libraries, suggesting our overall results may be robust against changes to $\gamma_i$. However, there are two differences with v3. First, $\PA_1$ seems to take longer to converge to the black hole's approaching limb in v3 as we go from zero-spin to strongly retrograde. This is expected; the lower adiabatic index drives higher midplane temperatures, which enables more pressure support and faster rotation \citep[see Appendix A of ][]{Conroy_2025}. Thus Doppler beaming plays a stronger role on the side of the receding limb than the approaching limb in v3 than in v5. 
Second, v3 seems to exhibit a wider standard deviation in $\PA_1$. 
Differences aside, the results are broadly consistent.

\section{$\PA_1$ Data from the Model Library}
\label{subsec:appendix_data}
Table \ref{tab:gauss_fits} provides summary fits to the $\PA_1$ distributions across models, marginalizing across $\rhigh$ and $\rlow$. The full set of fits are available on the Illinois Data Bank \citep{Conroy_2026rawdata}.
\begin{deluxetable}{ccccccccc}
    \tablecaption{Circular Gaussian Fits to the Brightness Asymmetry Position Angle $\PA_1$\label{tab:gauss_fits}}
    \tablewidth{\textwidth}
    \tablehead{
    \colhead{Source} & \colhead{MAD/SANE} & \colhead{Spin} & \colhead{inclination [deg]} & \colhead{$\rhigh$} & \colhead{$\rlow$} & \colhead{$\mu$ [deg]} & \colhead{$\sigma$ [deg]} & \colhead{$Mo$ [deg]} }
    \startdata
    M87* & MAD & $ -0.97 $ & $ 17 $ &  All  &  All  & $ 269 $ & $ 20 $ & $ 270 $ \\
    M87* & MAD & $ -0.9375 $ & $ 17 $ &  All  &  All  & $ 272 $ & $ 19 $ & $ 270 $ \\
    M87* & MAD & $ -0.85 $ & $ 17 $ &  All  &  All  & $ 271 $ & $ 27 $ & $ 280 $ \\
    M87* & MAD & $ -0.75 $ & $ 17 $ &  All  &  All  & $ 277 $ & $ 36 $ & $ 270 $ \\
    M87* & MAD & $ -0.5 $ & $ 17 $ &  All  &  All  & $ 284 $ & $ 54 $ & $ 270 $ \\
    M87* & MAD & $ -0.25 $ & $ 17 $ &  All  &  All  & $ 315 $ & $ 64 $ & $ 300 $ \\
    M87* & MAD & $ 0.0 $ & $ 163 $ &  All  &  All  & $ 144 $ & $ 68 $ & $ 140 $ \\
    M87* & MAD & $ 0.25 $ & $ 163 $ &  All  &  All  & $ 109 $ & $ 38 $ & $ 110 $ \\
    M87* & MAD & $ 0.5 $ & $ 163 $ &  All  &  All  & $ 95 $ & $ 29 $ & $ 90 $ \\
    M87* & MAD & $ 0.75 $ & $ 163 $ &  All  &  All  & $ 88 $ & $ 28 $ & $ 90 $ \\
    M87* & MAD & $ 0.85 $ & $ 163 $ &  All  &  All  & $ 90 $ & $ 24 $ & $ 90 $ \\
    M87* & MAD & $ 0.9375 $ & $ 163 $ &  All  &  All  & $ 83 $ & $ 29 $ & $ 90 $ \\
    M87* & MAD & $ 0.97 $ & $ 163 $ &  All  &  All  & $ 84 $ & $ 22 $ & $ 80 $ \\
    Sgr A* & MAD & $ -0.97 $ & $ 30 $ &  All  &  All  & $ 267 $ & $ 23 $ & $ 270 $ \\
    Sgr A* & MAD & $ -0.97 $ & $ 50 $ &  All  &  All  & $ 270 $ & $ 16 $ & $ 270 $ \\
    Sgr A* & MAD & $ -0.97 $ & $ 90 $ &  All  &  All  & $ 266 $ & $ 17 $ & $ 270 $ \\
    Sgr A* & MAD & $ -0.9375 $ & $ 30 $ &  All  &  All  & $ 266 $ & $ 23 $ & $ 280 $ \\
    Sgr A* & MAD & $ -0.9375 $ & $ 50 $ &  All  &  All  & $ 270 $ & $ 12 $ & $ 270 $ \\
    Sgr A* & MAD & $ -0.9375 $ & $ 90 $ &  All  &  All  & $ 268 $ & $ 12 $ & $ 270 $ \\
    Sgr A* & MAD & $ -0.85 $ & $ 30 $ &  All  &  All  & $ 264 $ & $ 26 $ & $ 270 $ \\
    Sgr A* & MAD & $ -0.85 $ & $ 50 $ &  All  &  All  & $ 268 $ & $ 18 $ & $ 270 $ \\
    Sgr A* & MAD & $ -0.85 $ & $ 90 $ &  All  &  All  & $ 267 $ & $ 15 $ & $ 270 $ \\
    Sgr A* & MAD & $ -0.75 $ & $ 30 $ &  All  &  All  & $ 276 $ & $ 40 $ & $ 280 $ \\
    Sgr A* & MAD & $ -0.75 $ & $ 50 $ &  All  &  All  & $ 272 $ & $ 25 $ & $ 270 $ \\
    Sgr A* & MAD & $ -0.75 $ & $ 90 $ &  All  &  All  & $ 265 $ & $ 19 $ & $ 270 $ \\
    Sgr A* & MAD & $ -0.5 $ & $ 30 $ &  All  &  All  & $ 286 $ & $ 51 $ & $ 290 $ \\
    Sgr A* & MAD & $ -0.5 $ & $ 50 $ &  All  &  All  & $ 284 $ & $ 36 $ & $ 300 $ \\
    Sgr A* & MAD & $ -0.5 $ & $ 90 $ &  All  &  All  & $ 264 $ & $ 35 $ & $ 270 $ \\
    Sgr A* & MAD & $ -0.25 $ & $ 30 $ &  All  &  All  & $ 355 $ & $ 58 $ & $ 350 $ \\
    Sgr A* & MAD & $ -0.25 $ & $ 50 $ &  All  &  All  & $ 351 $ & $ 48 $ & $ 330 $ \\
    Sgr A* & MAD & $ -0.25 $ & $ 90 $ &  All  &  All  & $ 300 $ & $ 103 $ & $ 270 $ \\
    Sgr A* & MAD & $ 0.0 $ & $ 30 $ &  All  &  All  & $ 36 $ & $ 48 $ & $ 50 $ \\
    Sgr A* & MAD & $ 0.0 $ & $ 50 $ &  All  &  All  & $ 36 $ & $ 41 $ & $ 40 $ \\
    Sgr A* & MAD & $ 0.0 $ & $ 90 $ &  All  &  All  & $ 90 $ & $ 45 $ & $ 90 $ \\
    Sgr A* & MAD & $ 0.25 $ & $ 30 $ &  All  &  All  & $ 81 $ & $ 32 $ & $ 70 $ \\
    Sgr A* & MAD & $ 0.25 $ & $ 50 $ &  All  &  All  & $ 83 $ & $ 23 $ & $ 70 $ \\
    Sgr A* & MAD & $ 0.25 $ & $ 90 $ &  All  &  All  & $ 90 $ & $ 13 $ & $ 90 $ \\
    Sgr A* & MAD & $ 0.5 $ & $ 30 $ &  All  &  All  & $ 93 $ & $ 28 $ & $ 90 $ \\
    Sgr A* & MAD & $ 0.5 $ & $ 50 $ &  All  &  All  & $ 92 $ & $ 16 $ & $ 90 $ \\
    Sgr A* & MAD & $ 0.5 $ & $ 90 $ &  All  &  All  & $ 90 $ & $ 9 $ & $ 90 $ \\
    Sgr A* & MAD & $ 0.75 $ & $ 30 $ &  All  &  All  & $ 108 $ & $ 32 $ & $ 100 $ \\
    Sgr A* & MAD & $ 0.75 $ & $ 50 $ &  All  &  All  & $ 99 $ & $ 17 $ & $ 100 $ \\
    Sgr A* & MAD & $ 0.75 $ & $ 90 $ &  All  &  All  & $ 89 $ & $ 13 $ & $ 90 $ \\
    Sgr A* & MAD & $ 0.85 $ & $ 30 $ &  All  &  All  & $ 102 $ & $ 30 $ & $ 90 $ \\
    Sgr A* & MAD & $ 0.85 $ & $ 50 $ &  All  &  All  & $ 97 $ & $ 16 $ & $ 90 $ \\
    Sgr A* & MAD & $ 0.85 $ & $ 90 $ &  All  &  All  & $ 91 $ & $ 14 $ & $ 90 $ \\
    Sgr A* & MAD & $ 0.9375 $ & $ 30 $ &  All  &  All  & $ 108 $ & $ 28 $ & $ 100 $ \\
    Sgr A* & MAD & $ 0.9375 $ & $ 50 $ &  All  &  All  & $ 99 $ & $ 18 $ & $ 100 $ \\
    Sgr A* & MAD & $ 0.9375 $ & $ 90 $ &  All  &  All  & $ 89 $ & $ 22 $ & $ 90 $ \\
    Sgr A* & MAD & $ 0.97 $ & $ 30 $ &  All  &  All  & $ 107 $ & $ 24 $ & $ 100 $ \\
    Sgr A* & MAD & $ 0.97 $ & $ 50 $ &  All  &  All  & $ 99 $ & $ 19 $ & $ 100 $ \\
    Sgr A* & MAD & $ 0.97 $ & $ 90 $ &  All  &  All  & $ 90 $ & $ 26 $ & $ 90 $ \\
    \enddata
    \tablecomments{Circular Gaussian fits to the $\PA_1$ distributions, with circular mean $\mu$, circular standard deviation $\sigma$, and mode $Mo$ (rounded to the nearest $10^\circ$), across model parameters. The full set of $\PA_1$ data in this work is available on the Illinois Data Bank: DOI 
    \href{https://doi.org/10.13012/B2IDB-7870444_V1}{10.13012/B2IDB-7870444\_V1}  \citep{Conroy_2026rawdata}. }
\end{deluxetable}

\end{appendix}
\bibliographystyle{aasjournal}
\bibliography{main}

\end{document}